# Demonstration of Sputtering Atomic Layer Augmented Deposition (*SALAD*): Aluminum Oxide-Copper Dielectric-Metal Nanocomposite Thin Films

Brian Giraldo[1], David M. Fryauf[1], Brandon Cheney[2], and Nobuhiko P. Kobayashi[1]

1. Nanostructured Energy Conversion Technology and Research (NECTAR), Department of Electrical and Computer Engineering, Baskin School of Engineering, University of California Santa Cruz, California 95064, United States

2. Department of Earth and Planetary Sciences, University of California Santa Cruz, California 95064, United States

**Abstract**:

Designing a thin film structure often begins with choosing a film deposition method that employs a specific primary process by which chemical species are formed and transported. In other words, a film deposition system in which two complementary deposition methods are hybridized should lead to new ways of designing thin film structures with structural complexity tailored at a level that has never been envisioned. This premise inspired us to uniquely combine atomic layer deposition (ALD) and magnetron sputtering (SPU) within a single deposition chamber; the combined film deposition system is referred to as sputtering atomic layer augmented deposition (*SALAD*). *SALAD* allows us to take full advantages of both ALD's precise and accurate precursor delivery and SPU's versatility in choices of chemical elements. A *SALAD* system was designed based on knowledge obtained by computational fluid dynamics with the goal of conceiving a film deposition system that satisfied deposition conditions distinctive for both ALD and SPU, and a prototype *SALAD* system was assembled by employing off-the-shelf vacuum components. As a demonstration, the *SALAD* system was utilized to deposit a unique nanocomposite made of a stack of aluminum oxide thin films by ALD and copper thin films by SPU – $AlO_x$-$Cu$ dielectric-metal (ACDM) nanocomposite thin films – on Si and glass substrates. Spectroscopic reflectivity collected on ACDM nanocomposite thin films shows unique dispersion features to which conventional effective medium theories used for describing optical properties of composites made of a dielectric host that contains metallic inclusions do not seem to simply apply.

## 1. Introduction

Searching for answers to questions at the boundary between different disciplines (e.g., magnetism and superconductivity) has repeatedly yielded discoveries of exotic phenomena related to quantum nature of materials. In such exploratory searches, the current approaches focus on uncovering hidden phenomena emerging from materials prepared by rather conventional synthesis techniques, in that, choices of constituent atomic elements set the level of distinctiveness of materials. However, in our view, *engineering* of synthesis techniques is overlooked, and we strongly believe that *engineering* breakthroughs in synthesis techniques will lead to materials possessing structural complexities that have never been envisioned, ultimately benefiting the *science* of materials – science enabled by engineering. For instance, sputtering (SPU) and atomic layer deposition (ALD) – two prevalent technologies for the deposition of thin films – have critical shortcomings. While SPU is capable of depositing thin films using flexible choices from a variety of chemical elements mixed at various compositions, it largely lacks accurate and precise control on delivering consistent amounts of chemical elements. In contrast, ALD allows deposition of thin films with accurate and precise control on thickness; however, it generally lacks explicit tunability of chemical composition. These shortcomings become noticeable when a thin film requires tight thickness control in the range of ~0.1 nm and chemical composition that needs to be varied continuously. In this paper, a new concept of hybrid thin film deposition system, based on SPU and ALD and referred to as Sputtering Atomic Layer Augmented Deposition (*SALAD*), is demonstrated. *SALAD* brings features not found in conventional stand-alone SPU or ALD, or in easily-imaginable cluster tools built upon SPU and ALD[1], allowing SPU and ALD to operate in a single environment in establishing, for instance, quasi-synchronous processes or sequential processes of the two deposition methods generally recognized as mutually incompatible from one another. *SALAD* allows us to perform "*in-situ*" tuning of surface composition at monolayer level without substantially altering a given or target thickness, which cannot be done by stand-alone SPU or ALD, or even by a cluster tool if the required total thickness is large (e.g., a few micrometers). A thin film structure that consists of aluminum oxide ($AlO_x$) and copper (Cu) is presented as a case exemplifying unique capabilities of *SALAD*. Two specific materials, $AlO_x$ and Cu, were chosen to illustrate distinctive capabilities of *SALAD* by emphasizing that *SALAD* enables the combination of various thin film materials generally difficult to be prepared by a single deposition method of either SPU or ALD.

## 2. Design of *SALAD*

Shown schematically in Fig. 1 is a cross sectional view of a *SALAD* system revealing its details with various critical dimensions. As shown in Fig. 1, the system consists of the following three main components: (1) SPU gun, (2) fast-acting spatial divider (FASD), and (3) precursor inlet/outlet for ALD. The three components are integrated on a single vacuum chamber with dimensions of the height $h_c$ and the diameter $w_c$ for a cylindrical chamber. The FASD plays a key role in uniquely establishing a deposition environment suitable for ALD within a deposition chamber optimized for SPU. The FASD also minimizes cross-contamination expected to occur when two deposition modes, ALD and SPU, run sequentially or concurrently. The FASD is inserted into or retracted from the deposition chamber within a time period of 1 s, without breaking the vacuum, to alternate two deposition environments – one for ALD and the other for SPU. A *SALAD* system was designed by carrying out 2D <u>c</u>omputational <u>f</u>luid <u>d</u>ynamics combined with <u>f</u>inite-<u>e</u>lement-<u>a</u>nalysis (CFD-FEA) to optimize, for instance, the width *w*, or diameter in case of a circular FASD, of the FASD in relation to the chamber width $w_c$ and the SPU throw distance *h*. Table 1 summarizes several key parameters used in the CFD-FEA. Since ALD and SPU co-exist in a single deposition chamber, it is critical to ensure that inserting the FASD creates a deposition environment suitable for ALD while *h* is optimized for SPU;

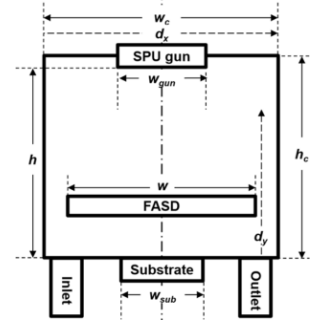

**Fig. 1:** A prototype *SALAD* system used for the demonstration is schematically shown, revealing its details and basic concept. The system is equipped with a sputtering cathode "SPU gun" and a fast-acting spatial divider "FASD". A substrate is placed underneath the FASD. "Inlet" and "Outlet" direct carrier gases and precursors used in ALD. Several dimensional parameters used in the CFD-FEA analysis are also denoted.

with the FASD inserted in the chamber, a laminar flow that contains ALD precursors over the substrate is established and maintained even though *h* is optimized for SPU. In Fig. 1, $d_x$ and $d_y$ represent two orthogonal directions along which distance from their respective origin is measured; in other words, $d_x$ = 30 mm denotes the point located 30 mm from the left wall of the vacuum chamber while the FASD is located at $d_y$ = 30 mm illustrates that the FASD is placed 30 mm above the substrate. For the CFD-FEA design, a rather small SPU gun was used ($w_{gun}$ = 26 mm), thus, it was important to ensure that optimum uniformity in the spatial density of SPU species ($\rho_{spu}$) coming from the SPU gun was achieved across the substrate.

| Parameter | Value |
|---|---|
| Chamber width ($w_c$) | 70 (mm) (The minimum linear size to accommodate a $\phi$1" diameter SPU gun) |
| Chamber height ($h_c$) | $h$ + 1 mm |
| Throw distance ($h$) | 10-120 mm (variable) |
| FASD thickness | 2 mm |
| FASD width ($w$) | 10-90 % of $w_c$ (variable) |
| FASD location | 10-30 mm above the substrate along the direction of $d_y$ (variable) |
| SPU gun width ($w_{gun}$) | 26 mm |
| Substrate width ($w_{sub}$) | 20 mm |
| PVD gas type, pressure | Gas type: argon at 300 K, Gas pressure: 1x$10^{-3}$ Torr |
| ALD gas inlet and outlet width | 5 mm |
| ALD inlet characteristics | Gas type: argon at 300 K, Inlet pressure: 760 Torr, Mass flow: 20 sccm |
| ALD outlet characteristics | Outlet pressure: 1x$10^{-3}$ Torr |

**Table 1:** Summary of the key parameters evaluated in the CFD-FEA analysis.

Fig. 2(a) and (b) illustrate two maps showing the distribution of the normalized $\rho_{spu}$ for two different $h$, (a) 10 mm and (b) 80 mm, indicating that $h$ significantly influences the distribution of $\rho_{spu}$ across the substrate, which is well known in designing SPU systems. Fig. 2(c) shows the dependence of nonuniformity defined as follows: ($\rho_{spu-max}$ − $\rho_{spu-min}$) / $\rho_{spu-min}$ where $\rho_{spu-max}$ and $\rho_{spu-min}$ are the maximum and the minimum $\rho_{spu}$ found across the substrate, clearly suggesting that $h$ needs to be at least 60 mm to minimize non-uniformity, although $h$ = 10 mm is expected to be more appropriate for ALD to maintain a laminar flow between the inlet and outlet. Fig. 3 shows the dependence of the distribution of the normalized density of gas molecules ($\rho_{ALD}$) used in ALD on w of the FASD. Gas molecules (i.e., ALD precursors) are injected into the chamber through the inlet and evacuated via the outlet assuming no reaction occurs over the substrate for three different $w$: (a) 7 mm, (b) 35 mm, and (c) 63 mm. The FASD is located at 10 mm above the substrate for all the three cases, indicating that a laminar flow is well established in Fig. 3(c). The establishment of a laminar flow is clearly seen in Fig. 4; it shows the dependence of the normalized $\rho_{ALD}$ on the distance from the substrate along the $d_y$ direction, as indicated in Fig. 3(a), originated at the center of the substrate. As indicated by the red arrow denoted by "Development of a laminar flow" in Fig. 4, a laminar flow develops progressively as $w$ increases and $w$ needs to be at least 90% of the chamber width $w_c$ to obtain substantial contrast in the normalized $\rho_{ALD}$ below and above the FASD for specific configurations used in the CFD-FEA analysis.

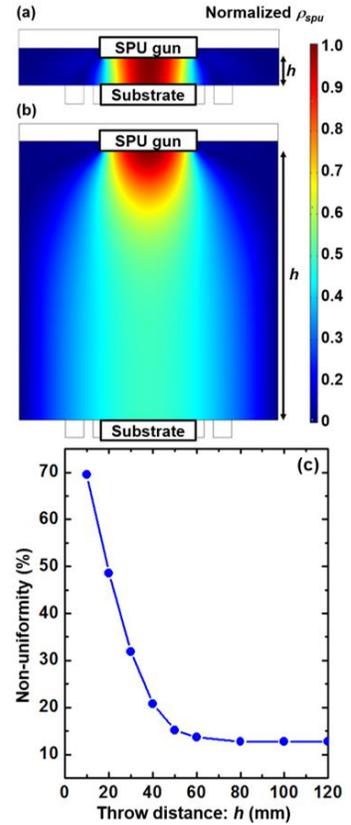

**Fig. 2:** (a)(b) Distribution maps of normalized $\rho_{spu}$ in the chamber with two different $h$, indicating the uniformity of $\rho_{spu}$, as well known, improves as $h$ increases. (c) Non-uniformity of $\rho_{spu}$ across the substrate plotted as a function of $h$, indicating $h$ needs to be at least 60 mm to minimize the non-uniformity.

Once the establishment of a laminar flow by inserting the FASD is visualized, it is worth evaluating the extent to which the FASD blocks SPU species moving towards the substrate. During a rather long ALD cycle (e.g., over 60 s), the SPU gun can be turned off in order to minimize the cross-contamination between SPU and ALD, however; for a short (e.g., ~3 s) ALD period, the SPU gun may need to remain turned on. Fig. 5 shows three maps displaying the distribution of normalized $\rho_{spu}$ in (a) the case in which no FASD is inserted and in (b) and (c) in which the FASD is placed at two different locations, 10 mm and 30 mm, from the substrate along the $d_y$ direction, respectively. While, without the FASD being inserted, SPU species generated by the SPU gun directly reach the substrate as seen in (a), the insertion of the FASD blocks the SPU species from directly reaching the substrate as indicated in both (b) and (c). However, a substantial portion of SPU species leaking through the gap between the FASD and the wall of the vacuum chamber reach the substrate as evident in comparing Fig. 5(b) and (c). Species reaching the substrate in this fashion also depends on the position of the FASD along the $d_y$ direction as further illustrated in Fig. 6. Fig. 6(a) and (b) show normalized $\rho_{spu}$ plotted as a function of distance along $d_x$ across the bottom of the vacuum chamber and along $d_x$ from the substrate as indicated in Fig. 5(c), respectively, indicating that the FASD with $w$ = 80 mm placed at 10 mm from the substrate (i.e., distance along $d_y$ = 10 mm) would effectively block

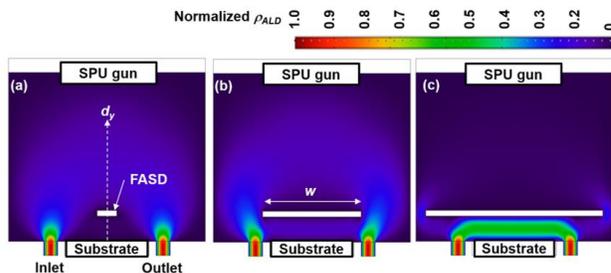

**Fig. 3:** The gas flow pattern underneath FASD is shown for various $w$: (a) $w$ = 7 mm, (b) $w$ = 35 mm, and (c) $w$ = 63 mm, respectively.

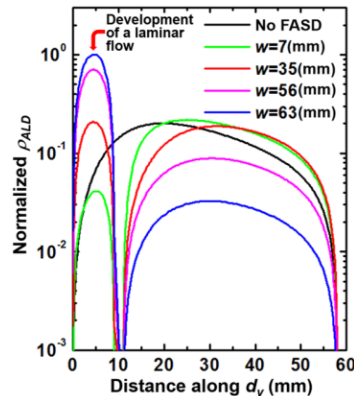

**Fig. 4:** The variation in $\rho_{ALD}$ along the $d_y$ direction denoted in Fig. 3(a), indicating the progressive development of a laminar flow as $w$ increases.

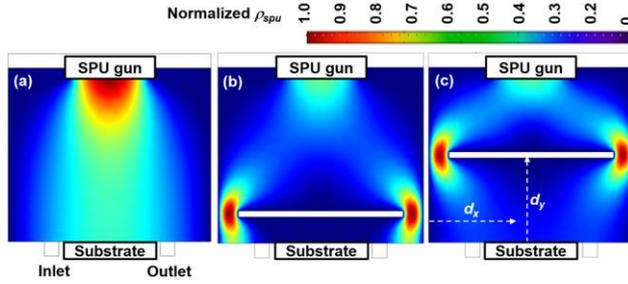

**Fig. 5:** Three maps display the distribution of $\rho_{spu}$ in SPU for (a) No FASD and two different locations of the FASD that blocks the line-of-sight path of SPU species going towards the substrate as shown in (b) and (c). In (b) and (c), the FASD is located at 10 mm and 30 mm from the substrate along the $d_y$ direction, respectively.

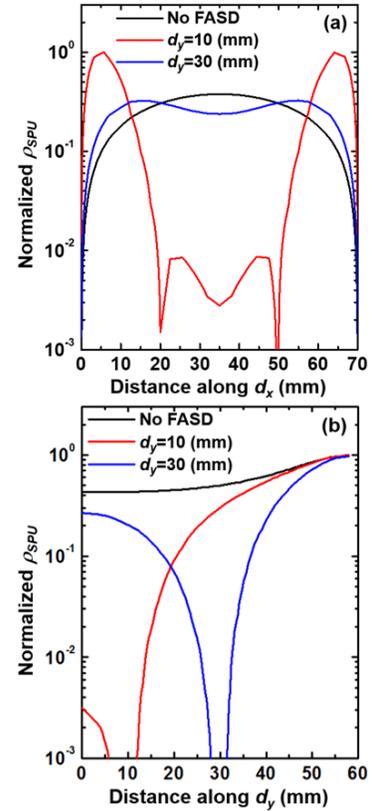

**Fig. 6:** Fig. 6(a) and (b) show normalized $\rho_{spu}$ plotted along the two different directions $d_x$ across the bottom of the chamber and $d_y$ above the substrate as indicated in Fig. 5(c), respectively.

SPU species while it creates a laminar flow. When the FASD is raised to 30 mm (i.e., distance along $d_y$ = 30 mm), SPU species appears to leak through the gap between the FASD and the wall of the vacuum chamber and reach the substrate, which makes the situation as unacceptable as that without the FASD (i.e., No FASD).

Although the CFD-FEA analysis was done on a *SALAD* system that accommodated a substrate with $w_{sub}$ = 20 mm as listed in Table 1, all the design parameters obtained in the analysis were scaled appropriately to assemble a *SALAD* system required for a series of experiments described below. The *SALAD* system used in the experiment was equipped with a SPU gun that accommodated a sputtering target with a diameter of 50 mm and was able to hold a substrate up to 100 mm in diameter.

## 3. Experimental

A series of deposition runs were carried out to distinctively demonstrate the capability of the *SALAD* system intended for unifying ALD and SPU in a single vacuum chamber. For the demonstration, a thin film structure made of aluminum oxide ($AlO_x$), dielectric material, and copper (Cu), metallic material, was used. $AlO_x$ and Cu were deposited by ALD and SPU, being referred to as "ALD-$AlO_x$" and "SPU-Cu", respectively. $AlO_x$ was chosen because of the level of its maturity as a dielectric material routinely deposited by ALD while Cu was chosen because of its free-electron density much higher than that of silver[2,3] and gold[4,5,6,7] – the two metallic materials commonly used as inclusions embedded in dielectric materials in order to explicitly tailor their optical properties as a whole. Aluminum (Al) could be an alternative to Cu; however, Al was avoided as it was expected to have formed $AlO_x$ when Al was exposed to the $AlO_x$ ALD process which used water as an oxidizer. In other words, $AlO_x$ resulting from the oxidation of Al deposited by SPU would not be easily distinguished from ALD-$AlO_x$. Cu chosen for the experiment also forms a variety of oxides including $Cu_2O$, $CuO$, and $Cu_4O_3$ that would significantly modify dielectric properties of ALD-$AlO_x$ by behaving as a band insulator[8], a charge transfer insulator[9], and a mixed-valence compound consisting of $Cu_2O$ and $CuO$[10], respectively. However, these oxides can be clearly distinguished from ALD-$AlO_x$ using analytical perspectives and are suitable for studying intermediate phases made by alternatively depositing ALD-$AlO_x$ and SPU-Cu being referred to as $\underline{AlO_x}$-$\underline{Cu}$ $\underline{d}$ielectric-$\underline{m}$etal (ACDM) nanocomposite thin films.

Table 2 summarizes the deposition parameters set for ALD and SPU in depositing the ACDM nanocomposite thin films. These specific deposition parameters were explicitly obtained through separately calibrating ALD and SPU processes in the *SALAD* system with the goal of achieving an appropriate deposition rate required for designing a single SPU step that delivered a specific amount of Cu with a corresponding equivalent-thickness in reference to the self-limited deposition rate indefinitely provided by the ALD process. In other words, the ALD process offered a self-limited deposition rate $\Gamma_{ALD}$ = 0.13 nm/cycle for ALD-$AlO_x$; thus the deposition rate of the SPU process $\Gamma_{SPU}$ for SPU-Cu was adjusted to be 0.04 nm/s so that the total amount of SPU-Cu was directly controlled by simply specifying a unique duration ($t_{SPU}$) of deposited SPU-Cu. A FilmTek 4000 spectroscopic reflectometry/ellipsometry (SRE) equipped with a UV-Visible light source was used to obtain $\Gamma_{ALD}$ and $\Gamma_{SPU}$. A SPU step with $t_{SPU}$ was inserted between two ALD cycles, which defined a single *SALAD* deposition cycle as illustrated in Fig. 7. In Fig. 7, the change in the background pressure in the *SALAD* system is plotted as a function of time revealing three *SALAD* cycles (note: only three out of the total 300 *SALAD* cycles are shown) recorded during the deposition of an ACDM nanocomposite thin film with $t_{SPU}$ = 3.0 s. A single *SALAD* cycle comprises: a Al(CH$_3$)$_3$ pulse indicated by a green vertical arrow, a purge for Al(CH$_3$)$_3$, a H$_2$O pulse marked by a red vertical arrow, a purge for H$_2$O, and a subsequent SPU step with a specific $t_{SPU}$ indicated by a red double-headed horizontal arrow. In Fig. 7, a unique feature in *SALAD* cycles is further revealed as follows: after the completion of a purge for H$_2$O, an intermediate evacuation step, observed as a sudden drop in the pressure indicated by a black vertical arrow, occurs before initiating a SPU step by temporarily stopping the flow of argon carrier gas for ALD-AlO$_x$ that creates a laminar flow along the direction perpendicular to the surface normal of the substrate. The laminar flow established for ALD was found to create an impenetrable barrier that stopped Cu species generated in a SPU step from reaching the substrate. In this inaugural experiment, a SPU step was inserted after a full ALD cycle was completed due to its simplicity that avoided an interruption of an ALD cycle. Needless to say, a SPU step can be inserted at any point (e.g., after a Al(CH$_3$)$_3$ pulse, after a purge for Al(CH$_3$)$_3$, after a H$_2$O pulse, or after a purge for H$_2$O) within a single ALD cycle.

A series of ACDM nanocomposite thin film samples was deposited on glass substrates (Corning 2497) by repeating a *SALAD* cycle comprising of an ALD step followed by a SPU step with a specific $t_{SPU}$ varied in the rage from 1.3 s to 7 s corresponding, given $\Gamma_{SPU}$ = 0.04 nm/s, to the amount of SPU-Cu equivalent to 0.05-0.28 nm thickness. The total number of *SALAD* cycles was fixed to 300 for all the samples. All the glass substrates were used as they were directly pulled from their original container, and no pre-deposition cleaning procedure was done. The following assumption is made to quantitatively distinguish the samples prepared with varied $t_{SPU}$: a single *SALAD* cycle yields a layer of ALD-AlO$_x$ with thickness of 0.13 nm followed by a layer of SPU-Cu with a nominal thickness of 0.04 nm/s x $t_{SPU}$ with an assumption that no intermixing of the two occurs. Although this assumption needs to be verified by carrying out detailed structural and chemical analysis of the samples, for the purpose of this paper, the ACDM nanocomposite thin film samples prepared with varied $t_{SPU}$ are contemplated as (AlO$_x$)$_n$(Cu)$_m$ where $n/m$ = ($\Gamma_{ALD}$ x 1 ALD cycle) / ($\Gamma_{SPU}$ x $t_{SPU}$) and $n + m$ = 1. By solving for $n$ and $m$ for a given $t_{SPU}$, the prepared seven variations of ACDM nanocomposite thin film samples are identified by using $m$ as listed in Table 3. In addition to the ACDM nanocomposite thin film samples, a 120 nm thick SPU-Cu (i.e., no ALD-AlO$_x$ was deposited) sample and a 30 nm thick ALD-AlO$_x$ (i.e., no SPU-Cu was deposited) sample were prepared as two reference samples.

| ALD parameters | |
|---|---|
| Aluminum precursor | Al(CH$_3$)$_3$ |
| Oxygen precursor | H$_2$O |
| Purge/carrier gas | Argon |
| Duration of an Al(CH$_3$)$_3$ pulse | 80 ms |
| Duration of a H$_2$O pulse | 150 ms |
| Purge duration after an Al(CH$_3$)$_3$ pulse | 10 s |
| Purge duration after a H$_2$O pulse | 10s |
| Purge/carrier gas flow rate | 20 sccm |
| Deposition temperature | 150 °C |
| Base pressure | 0.14 Torr |
| Deposition rate $G_{ALD}$ | 0.13 nm/cycle |

| SPU parameters | |
|---|---|
| Target material | Cu (purity 99.999%) |
| DC power density | 5 W cm$^{-2}$ |
| Sputtering gas | Argon |
| Deposition temperature | 150 °C |
| Deposition pressure | 0.24 Torr |
| Deposition rate $G_{SPU}$ | 0.04 nm/s |

**Table 2:** Summary of the ALD and SPU conditions used for depositing the AlO$_x$-Cu nanocomposite thin film structures.

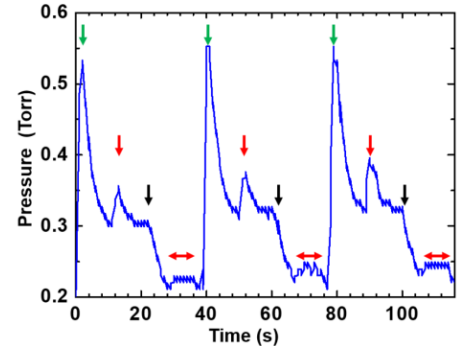

**Fig. 7:** Time-progressive change in the background pressure in the *SALAD* system shows three *SALAD* cycles recorded during the deposition of one of the AlO$_x$-Cu nanocomposite thin film structures used in the study. Each *SALAD* cycle comprises: a TMA pulse denoted by the green arrows, a purge for TMA, a H$_2$O pulse denoted by the blue arrows, a purge for H$_2$O, and a SPU duration indicated by the red horizontal double-head arrows.

| $t_{SPU}$ (s) | n | m |
|---|---|---|
| 1.3 | 0.71 | 0.29 |
| 2 | 0.62 | 0.38 |
| 3 | 0.52 | 0.48 |
| 4 | 0.45 | 0.55 |
| 5 | 0.39 | 0.61 |
| 6 | 0.35 | 0.65 |
| 7 | 0.32 | 0.68 |

**Table 3:** $n$ and $m$ calculated for various $t_{SPU}$ used in the experiment providing seven variations of ACDM nanocomposite thin film samples identified by $m$.

## 4. Results and Discussion

Energy dispersion spectroscopy (EDS) analysis of the ACDM nanocomposite thin film samples was carried out in a scanning electron microscope to obtain effective atomic composition of constituent chemical elements including Cu, oxygen (O), Al, and carbon (C) expected to be present in the samples. Specific conditions used in the EDS analysis are as follows: acceleration voltage of 15 kV and electron current of 1.6 nA. Shown in Fig. 8(a) is weight percent (wt%) of Cu (solid red line). Wt% of O (blue solid line), Al (orange solid line), and C (green solid line) are shown in (b) plotted as a function of $m$. The wt% of each of Cu, O, and Al – referred to as [Cu], [O], and [Al] – changes linearly as $m$ changes while wt% of C – referred to as [C] – remains almost constant or decreases slightly as $m$ increases. [Cu] monotonically increases while [Al] and [O] decreases as $m$ increases. The ratio of [Al] to [O] remains in the rage of 1.3-1.5 for $m$ = 0.29-0.68, reflecting the fact that x of ALD-AlO$_x$ is comparable or slightly smaller than 1.5 expected for stoichiometric Al$_2$O$_3$. Since ALD-AlO$_x$ cycles were fixed to 300 for all the samples, the increase in [C] in (a) and corresponding decreases in [Al] and [O] in (b) are a result of increase in [C] as $m$ increases. The trend of [C] slightly decreasing from 8 to 5 wt% as $m$ increases would be viewed as being consistent with the decrease in [Al] originated to ALD-AlO$_x$ made by using a precursor, trimethylaluminum, that contains C, as well as the fact that Al acting as a getter of such impurities as C present in the deposition environment. The EDS analysis clearly demonstrate *SALAD*'s capability of explicitly controlling [C] by SPU embedded in the ALD-AlO$_x$ host. The linear change of [Cu], [O], and [Al] as $m$ changes indicates that SPU and ALD processes done sequentially were independent. In other words, the two processes had minimum influence on one another.

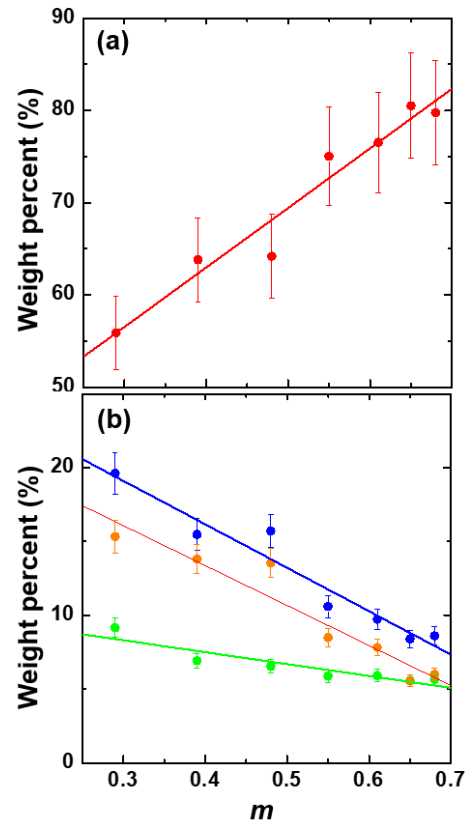

**Fig. 8:** Weight percent (wt%) of Cu in (a) and O (blue solid line), Al (orange solid line), and C (green solid line) in (b). The wt% of these chemical elements were obtained by using EDS spectra collected on the AlO$_x$-Cu nanocomposite thin film structures prepared with various $m$. The wt% is plotted as a function of $m$.

In this demonstration of *SALAD*, our focus was on optical properties – a mirror image of dielectric properties – of the ACDM nanocomposite thin film samples, which was expected to be modified by the inclusion of SPU-Cu in the samples. Measurements of reflectance spectra ($R$ spectra) were carried out to elucidate unique optical properties that would emerge from the samples characterized by their complexities of the way two materials, ALD-AlO$_x$ and SPU-Cu, are sequentially deposited by *SALAD*. Fig. 9(a)(b) shows $R$ spectra collected from the seven samples with various $m$ in the range of 0.29-0.68 as previously defined on Table 3. The $R$ spectra are presented in two different ranges of wavelength, 200-600 nm in (a) and 500-2500 nm in (b), to highlight the details of specific features appearing in these two spectral ranges. $R$ spectra obtained from the two reference samples – a 120 nm thick SPU-Cu ($m$ = 1) and a 30 nm thick ALD-AlO$_x$ ($m$ = 0) – are also displayed in Fig. 9(a)(b), showing $R$ dispersions generally expected for bulk forms of pure Cu and amorphous Al$_2$O$_3$. Fig. 9(a) clearly indicates that there exists a progressive development of $R$ in the range of 200-600 nm as $m$ increases. The wavelength range of 200-600 nm represents a range of wavelengths shorter than ~560 nm associated with the plasma reflection edge of bulk Cu. In other words, the features that appear in the $R$ spectra in Fig. 9(a) are expected to be dominated by the inter-band transitions rather than contributions associated with free-electrons. The absorption coefficient of bulk Cu measured in the wavelength range of 200-600 nm was found to exhibit features originated to three inter-band transitions at various symmetry points L2'-L1, X5-X4', and L3-Fermi surface[11]; thus the undulating features that appear in the wavelength range of 200-600 nm in Fig. 9(a) are most likely associated with the progressive formation of well-defined electronic bands of Cu, which can be further viewed as the increase in the total volume of Cu as $m$ increases by a fraction, highlighting precise and accurate control on delivering a specific amount of SPU-Cu physically separated by yet another precise and accurate control of an atomic layer of ALD-AlO$_x$, being made possible by *SALAD*. The statement *"the progressive formation of well-defined electronic bands of Cu"* needs to be further clarified and validated by performing such complementary characterization as Raman scattering spectroscopy and x-ray photoelectron spectroscopy to elaborate a physical and chemical picture of multiple Cu – continuous or non-continuous – layers, rather than bulk Cu, effectively separated by a single ALD-AlO$_x$ layer in the framework of the formation of electronic bands.

On the $R$ spectra in the range of wavelength from 600 to 2500 nm (i.e., the range of wavelength longer than the plasma reflection edge of bulk Cu) in Fig. 9(b), one would expect the dispersion of $R$ collected from the ACDM nanocomposite thin film samples to reflect characteristics of $R$ of the two constituent materials – Cu and AlO$_x$ – skillfully blended. Optical properties of bulk metals including Cu in this spectral range are often described by using the Drude-Lorentz theory in which free-electrons (i.e. electrons experiencing no restoring force) play the major role in determining complex dielectric constants (i.e., complex optical constants) relevant to the $R$ spectra. In the description of the Drude-Lorenz theory, the equation of motion for the displacement of a free-electron includes the frictional damping force $F_\gamma$ that accounts for the retarding action of the surrounding medium. In such metals as Cu, $F_\gamma$ is often set to zero to explain the $R$ spectrum of SPU-Cu ($m$ = 1) being close to 100%. The series of $R$ spectra for the different $m$ in Fig. 9(b) shows no distinct tendency in the way by which the spectra change as $m$ changes in contrast to the definitive changes in the $R$ spectra in Fig. 9(a). Interestingly, the sharp increase in the $R$ spectrum of SPU-Cu ($m$ = 1) seen at ~560 nm seems to emerge in none of the $R$ spectra collected from the samples with $m$ = 0.29-0.68. In a qualitative perspective, $F_\gamma$ appears to be a factor that should not be set to zero for the samples with $m$ = 0.29-0.68, suggesting that $F_\gamma$ is much larger in the ACDM nanocomposite thin film samples presumably because of the presence of ALD-AlO$_x$ layers even for $m$ = 0.69 for which the total weight percent of Cu is ~80% as seen in Fig. 8(a). It is also conceivable that the inter-band transition discussed earlier for the range of wavelength shorter than the plasma reflection edge of 560 nm somehow extends well above the plasma reflection edge, suggesting that ALD-AlO$_x$ layers may have given rise to additional inter-band transitions associated with SPU-Cu in the ACDM nanocomposite thin film samples because of, for instance, the alternations on the density of states in 3$d$ and 4$s$ bands of Cu.

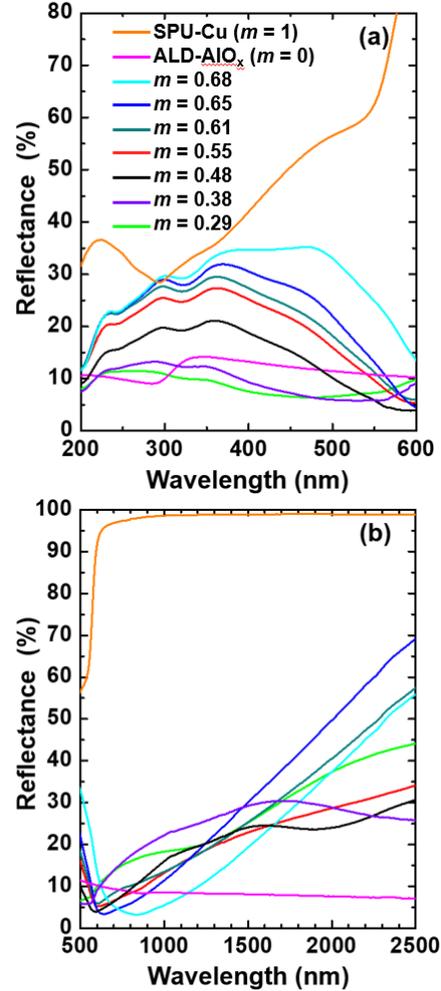

**Fig. 9:** Reflectance $R$ spectra of the ACDM nanocomposite thin film structures prepared with various $m$ along with those collected from SPU Cu ($m$ = 1) and ALD AlO$_x$ ($m$ = 0). The reflectance spectra are plotted over two different ranges: (a) 200-600 nm and (b) 500-2500 nm to highlight features distinctive in their respective ranges.

Contributions to the unique $R$ spectrum in Fig. 9(a) and (b) at the macroscopic-level rather than the microscopic-level should not be ruled out – for instance, optical interference occurring in a structure in which two types of layers having refractive index significantly different one another. However, since the optical thickness of each of ALD-AlO$_x$ and SPU-Cu layers in the ACDM nanocomposite thin film samples is much smaller than a quarter of the wavelengths of the light used in measuring the $R$ spectra, destructive and constructive interferences by which conventional distributed Bragg reflectors and band-pass filters operate are not relevant. The skin effect often used in discussing AC conductivity in metals would be another contributor to the peculiar $R$ spectrum illustrated in Fig. 9(a) and (b). For instance, the skin depth of Cu is estimated to be 2.7-5.9 nm using the conventional formula[12] for electromagnetic waves with wavelengths 500-2500 nm relevant to our experiment, respectively, which are up to two orders of magnitude larger than the thickness of a single SPU-Cu layer in the ACDM nanocomposite thin film samples, indicating that exploiting the concept of the skin effect is invalid. Another perspective, conventional nanocomposites A$_y$B$_{1-y}$ made of a dielectric host made of element A that contains metal inclusions made of element B often show complex optical characteristics when the dielectric-metal compositional ratio y is in the range of 0.2 - 0.8[13], which is often described within the framework of effective medium theories including the Maxwell Garnett approximation intended to undertake the interaction between electromagnetic wave and such mediums as a colloidal solution of metallic particles suspended in a dielectric host [14]. In the Maxwell Garnett approximation, a complex medium – effective medium – is assumed to exhibit effective permittivity approximated in terms of the permittivity and volume fractions of the independent constituents of the complex medium. This approximation, however, is clearly not valid in the ACDM nanocomposite thin film samples because of the presence of a structural periodicity (i.e., ALD-AlO$_x$ and SPU-Cu layers being stuck on top of each other) while the random distribution of metallic particles within a dielectric host is assumed in the Maxwell Garnett approximation. Further assessment is currently in progress to understand the unique *R* dispersion considering the structural periodicity introduced by *SALAD*.

## 5. Conclusion

Our vision of creating a unique deposition system capable of two deposition techniques deemed incompatible each other has been realized by demonstrating *SALAD* that has yielded the ACDM nanocomposite thin film samples made of ALD-AlO$_x$ and SPU-Cu with optical properties that could not be simply assessed by interpolating individual optical properties of its constituent materials. The key to engineering *SALAD* was the modeling of the gas flow in the chamber; our CFD-FEA allowed us to optimize chamber design parameters such as FASD width, chamber width, and SPU throw distance. Our demonstration would open a new frontier of materials science exploration by providing unique thin film structures made of a wide range of materials, which would not be easily realized by employing a thin film deposition system based on a single deposition method. We are confident that the extreme level of control over deposition rate and the enhanced ability to deliver a variety of chemical elements is producing materials with structural complexities that lead to novel physical phenomena.